\documentclass[12pt,letterpaper]{article}
\usepackage{amsmath,amssymb,array,calc,rotating,epsfig,psfrag,amscd, cite}

\makeatletter
\renewcommand\section{\@startsection {section}{1}{\z@}%
                                   {-3.5ex \@plus -1ex \@minus -.2ex}
                                   {2.3ex \@plus.2ex}%
                                   {\normalfont\large\bfseries}}
\renewcommand\subsection{\@startsection{subsection}{2}{\z@}%
                                     {-3.25ex\@plus -1ex \@minus -.2ex}%
                                     {1.5ex \@plus .2ex}%
                                     {\normalfont\bfseries}}
\makeatother

\let\non\nonumber

\let\a=\alpha\let\b=\beta\let\d=\delta

\let\s=\sigma

\let\S=\Sigma

\newcommand{\bea}{\begin{eqnarray}}
\newcommand{\eea}{\end{eqnarray}}
\newcommand{\be}{\begin{equation}}
\newcommand{\ee}{\end{equation}}

\newcommand{\p}{\partial}

\newcommand{\C}[1]{$(\ref{#1})$}

\def\IZ{\relax\ifmmode\mathchoice
{\hbox{\cmss Z\kern-.4em Z}}{\hbox{\cmss Z\kern-.4em Z}}
{\lower.9pt\hbox{\cmsss Z\kern-.4em Z}} {\lower1.2pt\hbox{\cmsss
Z\kern-.4em Z}}\else{\cmss Z\kern-.4em Z}\fi}
\def\IR{\relax{\rm I\kern-.18em R}}

\def\one{{\hbox{ 1\kern-.8mm l}}}

\newlength{\bredde}
\def\slash#1{\settowidth{\bredde}{$#1$}\ifmmode\,\raisebox{.15ex}{/}
\hspace*{-\bredde} #1\else$\,\raisebox{.15ex}{/}\hspace*{-\bredde}
#1$\fi}

\newsavebox{\zzzbar}
\sbox{\zzzbar}
  {\setlength{\unitlength}{0.9em}
  \begin{picture}(0.6,0.7)
  \thinlines
  \put(0,0){\line(1,0){0.6}}
  \put(0,0.75){\line(1,0){0.575}}
  \multiput(0,0)(0.0125,0.025){30}{\rule{0.3pt}{0.3pt}}
  \multiput(0.2,0)(0.0125,0.025){30}{\rule{0.3pt}{0.3pt}}
  \put(0,0.75){\line(0,-1){0.15}}
  \put(0.015,0.75){\line(0,-1){0.1}}
  \put(0.03,0.75){\line(0,-1){0.075}}
  \put(0.045,0.75){\line(0,-1){0.05}}
  \put(0.05,0.75){\line(0,-1){0.025}}
  \put(0.6,0){\line(0,1){0.15}}
  \put(0.585,0){\line(0,1){0.1}}
  \put(0.57,0){\line(0,1){0.075}}
  \put(0.555,0){\line(0,1){0.05}}
  \put(0.55,0){\line(0,1){0.025}}
  \end{picture}}

\newcommand{\ena}{\end{eqnarray}}
\newcommand{\beqa}{\begin{eqnarray}}
\newcommand{\eeqa}{\end{eqnarray}}

\renewcommand{\b}{\beta}
\newcommand{\g}{\gamma}

\def\a{\alpha}
\def\b{\beta}

\def\d{\delta}

\def\g{\gamma}

\def\s{\sigma}

\def\S{\Sigma}

\begin{document}
\begin{titlepage}

\begin{center}



\vskip 2 cm
{\Large \bf Supergravity limit of genus two modular graph functions in the worldline formalism}\\
\vskip 1.25 cm { Anirban Basu\footnote{email address:
    anirbanbasu@hri.res.in} } \\
{\vskip 0.5cm  Harish--Chandra Research Institute, HBNI, Chhatnag Road, Jhusi,\\
Allahabad 211019, India}

\end{center}

\vskip 2 cm

\begin{abstract}
\baselineskip=18pt

We consider the contributions up to the $D^{10}\mathcal{R}^4$ terms in the low momentum expansion of the two loop four graviton amplitude in maximal supergravity that arise in the field theory limit of genus two modular graph functions that result from the low momentum expansion of the four graviton amplitude in toroidally compactified type II string theory, using the worldline formalism of the first quantized superparticle. The expression for the two loop supergravity amplitude in the worldline formalism allows us to obtain contributions from the individual graphs, unlike the expression for the same amplitude obtained using unitarity cuts which only gives the total contribution from the sum of all the graphs. Our two loop analysis is field theoretic, and does not make explicit use of the genus two string amplitude.

\end{abstract}

\end{titlepage}


\section{Introduction}

The field theory limit of amplitudes in string theory yields useful information about the low energy behavior of the theory. In this note, our aim is to obtain information about the contributions in the field theory limit from genus two $Sp(4,\mathbb{Z})$ invariant modular graph functions to certain terms in the low momentum expansion of the two loop four graviton amplitude in maximal supergravity, using as little input as possible from string theory. 

 The terms in the low momentum expansion of the two loop four graviton amplitude up to the $D^{12}\mathcal{R}^4$ term have been obtained in~\cite{Green:1999pu,Green:2005ba,Green:2008bf}. To obtain the final expressions, one has to integrate over the skeleton diagram at two loops, which involves integrating over the three Schwinger parameters representing the lengths of the links of the skeleton diagram (the ultraviolet divergences have to be regularized by adding suitable counterterms). This is greatly facilitated by expressing this integral as the integral over the volume and complex structure modulus of an auxiliary torus. Then the integrand, at each order in the $\alpha'$ expansion, is decomposed into a sum of contributions each of which satisfies an eigenvalue equation when acted on by the $SL(2,\mathbb{Z})$ invariant Laplacian of the complex structure of the auxiliary torus. The integrand for the two loop amplitude has been obtained using unitarity cut techniques in~\cite{Bern:1998ug}, which is central to these calculations. This calculation must reproduce that obtained from the field theory limit of the genus two four graviton amplitude in toroidally compactified type II theory. Thus at every order in the $\alpha'$ expansion, the supergravity integrand of the $D^{2k}\mathcal{R}^4$ term is given by the sum of field theory limits of $Sp(4,\mathbb{Z})$ invariant graphs at genus two with distinct topology which result from the low momentum expansion of the string amplitude. The links of these modular graphs correspond to scalar Green functions on the worldsheet, while the vertices are the positions of insertions of the vertex operators on the worldsheet. 

Now the integrand  in~\cite{Bern:1998ug} gives the total contribution from all the graphs put together, and there is no way to obtain the contribution from the field theory limit of each individual graph by itself. It is interesting in its own right to analyze the contributions from these graphs individually to the field theory amplitude. To obtain these individual contributions, one can of course consider the graphs in the low momentum expansion of the genus two amplitude and take the field theory limit. However, we want to proceed differently, and answer the question in the context of field theory directly, making as little use of the string amplitude as possible. Thus we would like to perform these calculations in a formalism which mimics the structure of string amplitudes as closely as possible. We shall use the worldline formalism of the first quantized superparticle~\cite{Dai:2006vj,Green:2008bf}, where the final structure of the amplitude has ingredients which can be directly interpreted as coming from the field theory limit of the string amplitude.

It seems that the worldline formalism might be very useful in understanding the structure of these amplitudes at higher loops, and the calculations we present are only the simplest applications. This is because this formalism includes the contributions from the planar and non--planar diagrams automatically in a single compact expression, much like the single diagram in the string amplitude, and hence all relative numerical factors are fixed at one go. Also given that it reproduces the field theory limit of individual graphs, a knowledge of these amplitudes might be useful in understanding the multi--loop string integrands themselves, where very little is known beyond genus two. These $Sp(2g,\mathbb{Z})$ invariant modular graph functions at genus $g$ are expected to satisfy rich relations among themselves in spite of being topologically distinct. Some of their properties have been analyzed at genus one~\cite{D'Hoker:2015foa,Basu:2015ayg,DHoker:2015wxz,D'Hoker:2016jac,Basu:2016kli,Kleinschmidt:2017ege,Basu:2017nhs,Broedel:2018izr}, and they have been shown to satisfy eigenvalue equations. The worldline formalism captures their details in the field theory limit, and hence is potentially useful in understanding them.

Keeping this in mind, our aim will be to obtain the field theory contributions of individual $Sp(4,\mathbb{Z})$ invariant modular graphs to the supergravity amplitude up to the $D^{10}\mathcal{R}^4$ interaction, which can be generalized to all orders in the $\alpha'$ expansion. We shall do so using the worldline formalism. Needless to say, the various contributions add up to yield the results obtained in~\cite{Green:2008bf}.

It will be interesting to understand the worldline formalism at higher loops, and see how it constrains the string amplitude. The measure in the integrand of the amplitude is expected to be involved at higher loops as well as in theories with lesser supersymmetry, and hence determining it will be central to these calculations. This will involve generalizing the analysis, for example, in~\cite{Schmidt:1994zj,Schmidt:1994aq,Schubert:2001he}. This formalism should also provide insight into extending the soft theorems beyond tree level as discussed in~\cite{Sen:2017xjn}.             

We begin by giving relevant details of the two loop four graviton amplitude. We first consider the expression obtained by the unitarity cut technique, followed by the expression in the worldline formalism. We next consider the contributions of the individual graphs to terms up to the $D^{10}\mathcal{R}^4$ interaction. Then we show how these results match with the structure of the string amplitude. Finally, expressing the integrands in terms of the volume and complex structure of the auxiliary torus, we obtain eigenvalue equations satisfied by the graphs up to the $D^8\mathcal{R}^4$ interaction, the structure of which mimics that in~\cite{Green:2008bf}.    

It is important to note that our field theory analysis uses the Green function that does not descend from a conformally invariant Green function in the string amplitude. Thus though our expressions descend from modular invariant graphs, extra contributions arise in the field theory limit from modular invariant graphs which are conformally invariant as well. They are obtained by beginning with the conformally invariant Arakelov Green function, as described in~\cite{DHoker:2018mys}. It is only the sum total of all the contributions that is independent of the choice of the Green function. 

\section{The two loop supergravity amplitude from unitarity cuts}

In this section and the next, we give only the relevant details of the two loop four graviton amplitude that are needed for our purposes, and refer the reader to the literature for more details.    

The four graviton amplitude at two loops in type II supergravity compactified on $\mathbb{R}^{9-n,1}\times T^n$ can be determined using unitarity cuts, and hence avoids tedious Feynman diagram calculations. The expression is given by~\cite{Bern:1998ug} 
\be \label{E1}\mathcal{A} =  \mathcal{I} (S,T,U)\mathcal{R}^4,\ee
where we have dropped an overall irrelevant factor and set $\alpha' =4$. The Mandelstam variables are given by $S = -  g^{MN} (p_1 + p_2)_M (p_1 + p_2)_N, T = - g^{MN} (p_1 + p_4)_M (p_1 + p_4)_N, U= - g^{MN} (p_1 + p_3)_M (p_1 + p_3)_N$ where $g_{MN}$ is the flat metric on $\mathbb{R}^{9-n,1}$. Here $p_i^2 = 0$ for $i=1,\ldots, 4$ and $\sum_{i=1}^4 p_i^M = 0$.

The factor $\mathcal{I} (S,T,U)$ in \C{E1} is given by
\bea \label{E2}\mathcal{I} (S,T,U) &=&  S^2 \Big(I_P (S;T,U) + I_P (S;U,T) + I_{NP} (S;T,U) + I_{NP} (S;U,T)\Big)\non \\ &&+T^2 \Big(I_P (T;S,U) + I_P (T;U,S) + I_{NP} (T;S,U) + I_{NP} (T;U,S)\Big) \non \\ &&+ U^2 \Big(I_P (U;S,T) + I_P (U;T,S) + I_{NP} (U;S,T) + I_{NP} (U;T,S)\Big). \eea 
Now the S--channel planar contribution depicted by figure 1, is given by
\bea \label{P}I_P (S;T,U) = \frac{\pi^{10-n}}{\mathcal{V}_n^2}\int_0^\infty d L_1 d L_2 d L_3 \frac{\Gamma_{(n,n)}}{\Delta^{(10-n)/2}}  \int_0^{L_3} dt_3 \int_0^{t_3} d t_4 \int_0^{L_1} dt_2 \int_0^{t_2} dt_1 e^{h_P},\eea
where
\bea h_P &=&  \frac{1}{\Delta} \Big[-S\Big(t_1t_2(L_2+L_3) + t_3 t_4 (L_1+L_2)\Big) + TL_2(t_1 t_4 + t_2 t_3) + U L_2 (t_1 t_3 + t_2 t_4)\Big]\non \\ && + S(t_1 + t_4).\eea
Note that $I_P(S;T,U) \neq I_P(S;U,T)$.

Also the S--channel non--planar contribution depicted by figure 1, is given by
\bea \label{NP}I_{NP} (S;T,U) = \frac{\pi^{10-n}}{\mathcal{V}_n^2}\int_0^\infty d L_1  d L_2 d L_3 \frac{\Gamma_{(n,n)}}{\Delta^{(10-n)/2}} \int_0^{L_3} d t_3 \int_0^{L_2}dt_4  \int_0^{L_1}dt_2 \int_0^{t_2} dt_1 e^{h_{NP}}, \eea
where
\be h_{NP} = S t_1 + \frac{1}{\Delta} \Big[ S \Big( - t_1 t_2 (L_2 + L_3) + t_3 t_4 L_1\Big) + T(t_1 t_4 L_3 + t_2 t_3 L_2) + U(t_1 t_3 L_2 + t_2 t_4 L_3)\Big].\ee
Note that $I_{NP}(S;T,U) = I_{NP}(S;U,T).$ The remaining contributions in \C{E2} are obtained simply by permuting the external legs. 

\begin{figure}[ht]
\begin{center}
\[
\mbox{\begin{picture}(270,110)(0,0)
\includegraphics[scale=.6]{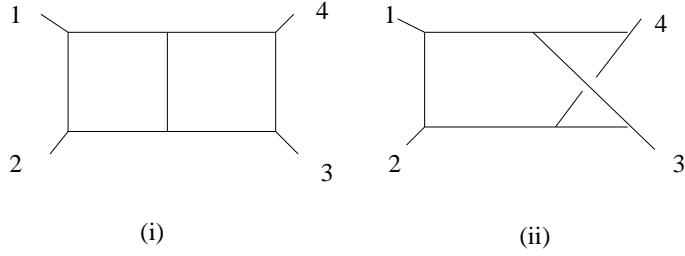}
\end{picture}}
\]
\caption{(i) planar, and (ii) non--planar diagrams}
\end{center}
\end{figure}

In the above expressions, we have that
\be \Delta = L_1 L_2  + L_2 L_3 + L_3 L_1,\ee
while the lattice factor is given by
\be \Gamma_{(n,n)} = \sum_{m_I ,n_I \in \mathbb{Z}} e^{-G^{IJ} \Big(L_1 m_I m_J + L_2 n_I n_J + L_3 (m+n)_I (m+n)_J\Big)}\ee
which involves a sum over the Kaluza--Klein momentum modes.
Here $G_{IJ}$ is the metric on $T^n$ with dimensionless volume $\mathcal{V}_n = \sqrt{{\rm det} G_{IJ}}$ in units of the string length. 

Defining
\be \s_n = S^n + T^n + U^n, \ee
we obtain expressions for the various interactions in the low momentum expansion. For the $D^4\mathcal{R}^4$ term we have that
\be \label{D4R4}\mathcal{A}_{D^4\mathcal{R}^4} = \frac{\pi^{10-n}\s_2\mathcal{R}^4}{6 \mathcal{V}_n^2} \int_0^\infty dL_1 dL_2 dL_3  \frac{\Gamma_{(n,n)}}{\Delta^{(6-n)/2}},\ee
while for the $D^6\mathcal{R}^4$ term we have that
\be \label{D6R4}\mathcal{A}_{D^6\mathcal{R}^4} = \frac{\pi^{10-n}\s_3\mathcal{R}^4}{72 \mathcal{V}_n^2} \int_0^\infty dL_1 dL_2 dL_3  \frac{\Gamma_{(n,n)}}{\Delta^{(6-n)/2}} \Big[ \mathcal{S} - \frac{5 \mathcal{M}}{\Delta}\Big] ,\ee
and for the $D^8\mathcal{R}^4$ term we have that
\bea \label{D8R4}\mathcal{A}_{D^8\mathcal{R}^4} = \frac{\pi^{10-n}\s_2^2 \mathcal{R}^4}{8640  \mathcal{V}_n^2} \int_0^\infty dL_1 dL_2 dL_3 \frac{\Gamma_{(n,n)}}{\Delta^{(6-n)/2}} \Big[4\mathcal{S}^2 -3 \Delta   - \frac{22\mathcal{M}\cal{S}}{\Delta} +\frac{32\mathcal{M}^2}{\Delta^2}\Big] .\eea
Finally for the $D^{10} \mathcal{R}^4$ term we have that
\bea \label{D10R4}\mathcal{A}_{D^{10}\mathcal{R}^4} &=& \frac{\pi^{10-n}\s_2 \s_3\mathcal{R}^4}{1088640  \mathcal{V}_n^2} \int_0^\infty dL_1 dL_2 dL_3   \frac{\Gamma_{(n,n)}}{\Delta^{(6-n)/2}}  \Big[45 {\cal{S}}^3+ 250 \mathcal{M} +  \frac{347\mathcal{M}^2 \cal{S}}{\Delta^2} \non \\ &&- 65 {\cal{S}}\Delta - \frac{145\mathcal{M}^3}{\Delta^3} - \frac{285\mathcal{M}{\cal{S}}^2}{\Delta}\Big] ,\eea
where we have defined
\be  \mathcal{S} = L_1+L_2+L_3, \quad \mathcal{M} = L_1 L_2 L_3. \ee
Note that $\Delta$, ${\cal{S}}$ and $\mathcal{M}$ are symmetric under interchange of $L_i$.

In obtaining the above expressions it is often useful to simplify intermediate expressions using identities like
\bea L_1^{10} + L_2^{10} +L_3^{10} &=& {\cal{S}}^{10} + 10  {\cal{S}}^7 \mathcal{M}  + 25  {\cal{S}}^4 \mathcal{M}^2  + 10 {\cal{S}} \mathcal{M}^3 -10 {\cal{S}}^8 \Delta - 60 {\cal{S}}^5 \mathcal{M} \Delta \non \\ && - 60  {\cal{S}}^2 \mathcal{M}^2 \Delta  + 35 {\cal{S}}^6 \Delta^2   + 100 {\cal{S}}^3 \mathcal{M} \Delta^2 + 15 \mathcal{M}^2 \Delta^2 -50 {\cal{S}}^4 \Delta^3  \non \\ && - 40 {\cal{S}} \mathcal{M} \Delta^3  + 25 {\cal{S}}^2 \Delta^4 - 2 \Delta^5  \eea
and express all quantities only in terms of $\Delta$, ${\cal{S}}$ and $\mathcal{M}$.  
 
\section{The two loop supergravity amplitude from the worldline formalism}

We now consider the same amplitude in the worldline formalism for the superparticle~\cite{Dai:2006vj,Green:2008bf}. It is given by
\be \label{E3}\mathcal{I} (S,T,U) = \frac{\pi^{10-n}}{2\mathcal{V}_n^2} \int_0^\infty d L_1 d L_2 d L_3 \Gamma_{(n,n)} \oint \prod_{r=1}^4 d t_r \frac{W^2}{\Delta^{(10-n)/2}}e^{-\sum_{r,s =1}^4 p_r \cdot p_s G_{rs}}.\ee
Here the various parameters in the integral have a geometric interpretation which matches directly with the expression given in the previous section, making the equivalence manifest as we explain below. The basic skeleton diagram for this amplitude is the two loop $\varphi^3$ vacuum bubble with three links and two vertices. Now $L_i$ are the lengths of the three links which are integrated over, and $t_r$ are the positions of insertion of the four vertex operators on the links. Thus $t_r$ is integrated over the skeleton diagram, while $G_{rs}$ is the Green function between vertices at points $t_r$ and $t_s$ on the graph.   

In order to write down the Green function $G_{rs}$, we define $ t_r \equiv t_r^{(k_r)} $ ($r=1,2,3,4$ and $k_r =1,2,3$)  which labels the position of insertion of the $r$--th vertex operator on the link $k_r$. Then $G_{rs}$  is defined by
\be \label{G}G_{rs} = -\frac{1}{2} d_{t_r^{(k_r)} t_s^{(k_s)}} + \frac{1}{2} (v^{(k_r)} - v^{(k_s)})_\a K^{-1}_{\a\b} (v^{(k_r)} - v^{(k_s)})_\b,\ee 
where $d_{t_r^{(k_r)} t_s^{(k_s)}}$ is the absolute value of the distance between points $t_r^{(k_r)}$ and $t_s^{(k_s)}$ on the graph\footnote{Thus we see that $G_{rr}=0$.}.  Here 
\be v_\a^{(k_r)} = t_r^{(k_r)} u_\a^{(k_r)},\ee
where $\alpha = 1,2$ for the two loops, and 
\be u^{(1)} = \left( \begin{array}{c} 1 \\ 0 \end{array}\right), \quad u^{(2)} = \left( \begin{array}{c} -1 \\ 1 \end{array}\right), \quad u^{(3)} = \left( \begin{array}{c} 0 \\ -1 \end{array}\right) \ee
are the column vectors corresponding to the three links.  
Also we have that
\be \label{inv}K^{-1} = \frac{1}{\Delta}\left( \begin{array}{cc} L_2 + L_3 & L_2\\ L_2 & L_1 + L_2\end{array}\right).\ee 
Thus if $t_r$ and $t_s$ are on the same line (thus $k_r = k_s$), then
\be G_{rs} = -\frac{1}{2} \vert t_r^{(k_r)} - t_s^{(k_r)} \vert +\frac{1}{2\Delta} (L_l+L_m) (t_r^{(k_r)} - t_s^{(k_r)})^2,\ee
where $l \neq m \neq k_r = k_s$. If they are on different lines ($k_r \neq k_s$) then
\be G_{rs} = -\frac{1}{2} (t_r^{(k_r)} + t_s^{(k_s)}) +\frac{\Big( (L_l+ L_{k_s}) (t_r^{(k_r)})^2 + (L_l + L_{k_r}) (t_s^{(k_s)})^2 + 2 t_r^{(k_r)} t_s^{(k_s)}L_l\Big)}{2\Delta}\ee
where $l \neq k_r \neq k_s$.

While this follows from the basic structure of the worldline formalism and is true in general, the measure factor $W^2$ is non--trivial and depends on the theory. For the maximally supersymmetric theories that we are considering, one can easily guess an expression for it at two loops by comparing with the structure obtained in \C{E1}. We see that $W^2$ vanishes if there are more than two insertions on the same link. The only non--vanishing possibilities are: $W^2 = S^2$ if $t_1, t_2$ (and/or $t_3,t_4$) are on the same link, $W^2 = T^2$ if $t_1, t_4$ (and/or $t_2,t_3$) are on the same link, and  $W^2 = U^2$ if $t_1, t_3$ (and/or $t_2,t_4$) are on the same link\footnote{In fact, $W$ is given by the expression
\be 3W = (T-U) \Delta_{12} \Delta_{34} + (S-T) \Delta_{13} \Delta_{42} + (U-S) \Delta_{14} \Delta_{23}\ee
where
\be \Delta_{rs} = \epsilon^{\a\b} u_\a^{(k_r)} u_\b^{(k_s)}\ee
for $\a,\b = 1,2$. This involves taking the field theory limit of the appropriate factor in the genus two string amplitude. }. We do not expect such simple expressions for the measure at higher loops, and the input from the string amplitude should play a crucial role in determining it\footnote{Determining the measure factor directly in field theory would be very interesting, at least in cases with maximal supersymmetry.}. However, the simplicity of the two loop amplitude in \C{E1} allows us to proceed without this input.       

Now using these results, in \C{E3} one has to integrate $t_r$ over the various links of the skeleton diagram keeping in mind the varying contributions of the integrand over different parts of moduli space which are determined by the nature of the Green function depending on whether the insertions are on the same or on different links. Thus analyzing the explicit $t_r$ dependence of the integral in \C{E3}, we see that it indeed reproduces \C{E1}. 

In fact, the planar and non--planar contributions given in figure 2 depict the expressions for $S^2 I_P (S;T,U)$ and $S^2 I_{NP} (S;T,U)$ respectively in the worldline formalism. This also gives a geometric meaning to the various parameters in \C{P} and \C{NP}.

\begin{figure}[ht]
\begin{center}
\[
\mbox{\begin{picture}(230,130)(0,0)
\includegraphics[scale=.65]{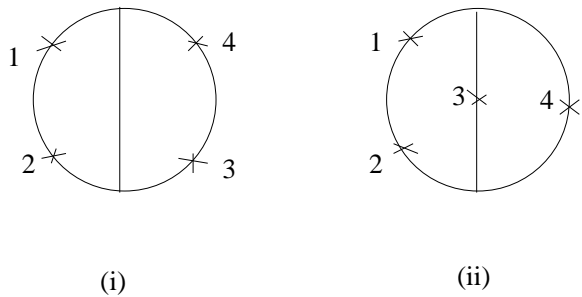}
\end{picture}}
\]
\caption{(i) planar, and (ii) non--planar diagrams}
\end{center}
\end{figure}

\section{The low momentum expansion of the two loop amplitude}

We now perform the low momentum expansion of the two loop supergravity amplitude up to the $D^{10} \mathcal{R}^4$ interaction. To do so, we use the expression \C{E3} rather than \C{E1} as this allows us to isolate contributions from the field theory limits of various modular graphs that arise in the string amplitude. This is because \C{E3} has a natural interpretation as the field theory limit of the string amplitude: the factor involving the exponential of the Green functions is the field theory limit of the Koba--Nielsen factor, and the momentum expansion of the amplitude which is an expansion in the Mandelstam variables is obtained by expanding the exponential factor in \C{E3}. This produces various topologically distinct graphs at various orders in this $\alpha'$ expansion where the links are given by the Green functions, and the vertices are given by the insertion points of the vertex operators on the worldline of the skeleton diagram. Thus these must be the field theory limits of the $Sp(4,\mathbb{Z})$ invariant modular graphs in the low momentum expansion of the string amplitude, hence producing a  one--to--one correspondence between the two sets of graphs at all orders in the $\alpha'$ expansion. Hence at two loops, the field theory amplitude in the worldline formalism completely fixes the contribution of the string theory graphs in the field theory limit, without actually explicitly needing to evaluate or even know the expression of the string amplitude.     

We now perform the low momentum expansion of \C{E3} to obtain the various terms  up to the $D^{10} \mathcal{R}^4$ term, and classify the contributions arising from topologically distinct graphs. This can be generalized to all orders in the $\alpha'$ expansion.

\subsection{The $D^4\mathcal{R}^4$  and $D^6\mathcal{R}^4$ terms}

For the $D^4\mathcal{R}^4$ term, since $W^2\sim S^2$, the exponential factor in \C{E3} is simply one, hence reproducing \C{D4R4}. This corresponds to the trivial graph with no links. For the $D^6\mathcal{R}^4$ term, a similar analysis reproduces \C{D6R4}. This corresponds to the only graph with one link given by figure 3 (we do not draw the vertices on the worldsheet which are not connected by Green functions, and they are implicit in the figures). Thus for these terms, either \C{E1} or \C{E3} is good enough for our purposes.

\begin{figure}[ht]
\begin{center}
\[
\mbox{\begin{picture}(110,20)(0,0)
\includegraphics[scale=1]{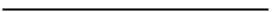}
\end{picture}}
\]
\caption{Graph for the $D^6\mathcal{R}^4$ term}
\end{center}
\end{figure}

Finally, in order to easily match with the structure of the string amplitude, we Poisson resum the sum over Kaluza--Klein momentum modes in the lattice sum to express it in terms of a sum over dual momentum modes. Thus the Poisson resummed lattice sum involves the metric $G_{IJ}$ (rather than the inverse metric $G^{IJ}$) which is what naturally arises in the string amplitude. Thus we get that 
\bea \mathcal{A}_{D^4\mathcal{R}^4} &=& \frac{\pi^{13} \s_2\mathcal{R}^4}{6\mathcal{V}_n^2} \int_0^\infty \frac{dL_1 dL_2 dL_3 }{\Delta^3} \hat{\Gamma}_{(n,n)}, \non \\ \mathcal{A}_{D^6\mathcal{R}^4} &=& \frac{\pi^{12} \s_3\mathcal{R}^4}{72\mathcal{V}_n^2} \int_0^\infty \frac{dL_1 dL_2 dL_3 }{\Delta^3} \hat{\Gamma}_{(n,n)} \Big({\cal{S}} - \frac{5{\mathcal{M}}}{\Delta}\Big),\eea
where the lattice sum is given by
\be \hat{\Gamma}_{(n,n)} = \mathcal{V}_n^2 \sum e^{-\pi G_{IJ} \Big(L_1 m_I m_J + L_2 n_I n_J + L_3 (m+n)_I (m+n)_J\Big)/\Delta}.\ee 

\subsection{The $D^8\mathcal{R}^4$ term}

From the $D^8\mathcal{R}^4$ term onwards in the low momentum expansion, we see the non--trivial role played by \C{E3} in isolating the contributions from the topologically distinct graphs. 

\begin{figure}[ht]
\begin{center}
\[
\mbox{\begin{picture}(250,120)(0,0)
\includegraphics[scale=.75]{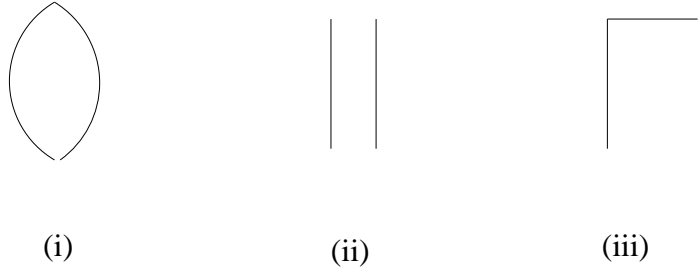}
\end{picture}}
\]
\caption{Graphs for the $D^8\mathcal{R}^4$ term}
\end{center}
\end{figure}

The $D^8\mathcal{R}^4$ term involves graphs with two links. These are the three topologically distinct graphs that arise from \C{E3} which are given in figure 4. For each graph, we write
\be \label{split}\mathcal{I} (S,T,U) = {\mathcal{I}}_P (S,T,U) + {\mathcal{I}}_{NP} (S,T,U)\ee
where ${\mathcal{I}}_P (S,T,U)$ and  ${\mathcal{I}}_{NP} (S,T,U)$ are the planar and non--planar contributions respectively\footnote{The total contribution from all the graphs is given by adding the expressions for the planar and non--planar integrals.}.  

To obtain these contributions, we define the integrals
\bea f_1^i &=& \int [dt]_i (2G_{12}^2 + 2G_{34}^2 + G_{14}^2 + G_{23}^2 +G_{13}^2 + G_{24}^2), \non \\ f_2^i&=& \int [dt]_i (2G_{12}G_{34} + G_{14} G_{23} + G_{13} G_{24}), \non \\ f_3^i &=&  \int [dt]_i (G_{12} + G_{34}) (G_{14} + G_{23} + G_{13} + G_{24}),\eea
which arise as the integrands of the graphs with distinct topologies. Here $i=P,NP$, and we have that
\bea \int [dt]_P &=& \int_0^{L_3} dt_3 \int_0^{t_3} d t_4 \int_0^{L_1} dt_2 \int_0^{t_2} dt_1, \non \\ \int [dt]_{NP} &=& \int_0^{L_3} d t_3 \int_0^{L_2}dt_4  \int_0^{L_1}dt_2 \int_0^{t_2} dt_1.\eea
Thus when $i=P$, the above integrals involve the planar measure with the Green functions given according to figure 2. Also when $i=NP$, the above integrals involve the non--planar measure with the Green functions given according to figure 2.

\subsubsection{The planar contributions}

We first consider the planar contributions ${\mathcal{I}}_P (S,T,U)$.
For the graphs (i), (ii) and (iii) in figure 4, we have that
\bea {\mathcal{I}}_P^{(1)} (S,T,U) &=& \frac{\pi^{10-n}\s_2^2}{4\mathcal{V}_n^2}  \int_0^\infty dL_1 dL_2 dL_3 \frac{\Gamma_{(n,n)}}{ \Delta^{(6-n)/2}} f_1^P , \non \\ {\mathcal{I}}_P^{(2)} (S,T,U) &=& \frac{\pi^{10-n}\s_2^2}{2\mathcal{V}_n^2}  \int_0^\infty dL_1 dL_2 dL_3 \frac{\Gamma_{(n,n)}}{ \Delta^{(6-n)/2}}f_2^P, \non \\  {\mathcal{I}}_P^{(3)} (S,T,U) &=& -\frac{\pi^{10-n}\s_2^2}{2\mathcal{V}_n^2} \int_0^\infty dL_1 dL_2 dL_3 \frac{\Gamma_{(n,n)}}{ \Delta^{(6-n)/2}}f_3^P\eea
respectively, where
\bea  f_1^P &=& \frac{1}{2160}\Big(9 {\cal{S}}^2 -8 \Delta +  \frac{20\mathcal{M} {\cal{S}}}{\Delta} -  \frac{48\mathcal{M}^3 {\cal{S}}}{\Delta^4}  -  \frac{28\mathcal{M}^2 {\cal{S}}^2}{\Delta^3} +  \frac{\mathcal{M}}{\Delta^2} (13 \mathcal{M} - 18 {\cal{S}}^3)\Big), \non \\ f_2^P &=& \frac{1}{864}\Big( {\cal{S}}^2 + \Delta - \frac{ 2\mathcal{M} {\cal{S}}}{\Delta} - \frac{4\mathcal{M}^3 {\cal{S}}}{\Delta^4}   - \frac{6\mathcal{M}^2 {\cal{S}}^2}{\Delta^3} + \frac{ \mathcal{M}}{\Delta^2}(5\mathcal{M} - 2{\cal{S}}^3)\Big) , \non \\ f_3^P &=& \frac{1}{2160}\Big( 5 {\cal{S}}^2 + \frac{4\mathcal{M} {\cal{S}}}{\Delta} - \frac{24\mathcal{M}^3 {\cal{S}}}{\Delta^4}   - \frac{ 28\mathcal{M}^2 {\cal{S}}^2}{\Delta^3} + \frac{\mathcal{M}}{\Delta^2} (13 \mathcal{M} - 10 {\cal{S}}^3)\Big).\eea

\subsubsection{The non--planar contributions}

We next consider the non--planar contributions. Again for the graphs (i), (ii) and (iii) from figure 4, we have that
\bea {\mathcal{I}}_{NP}^{(1)}(S,T,U) &=& \frac{\pi^{10-n}\s_2^2}{4\mathcal{V}_n^2}  \int_0^\infty dL_1 dL_2 dL_3 \frac{\Gamma_{(n,n)}}{ \Delta^{(6-n)/2}} f_1^{NP}, \non \\ {\mathcal{I}}_{NP}^{(2)} (S,T,U) &=& \frac{\pi^{10-n}\s_2^2}{2\mathcal{V}_n^2} \int_0^\infty dL_1 dL_2 dL_3 \frac{\Gamma_{(n,n)}}{ \Delta^{(6-n)/2}} f_2^{NP}, \non \\ {\mathcal{I}}_{NP}^{(3)} (S,T,U) &=& -\frac{\pi^{10-n}\s_2^2}{2\mathcal{V}_n^2}  \int_0^\infty dL_1 dL_2 dL_3 \frac{\Gamma_{(n,n)}}{ \Delta^{(6-n)/2}}f_3^{NP}\eea
respectively, where
\bea f_1^{NP} &=& \frac{\mathcal{M}}{1080\Delta}\Big(- {16\cal{S}} + \frac{24 \mathcal{M}^2 {\cal{S}}}{\Delta^3} + \frac{14  \mathcal{M} {\cal{S}}^2}{\Delta^2} -  \frac{3(\mathcal{M}-3{\cal{S}}^3)}{\Delta}\Big), \non \\  f_2^{NP} &=& \frac{\mathcal{M}}{432\Delta^4}(2\mathcal{M} + \Delta {\cal{S}})\Big(\mathcal{M} {\cal{S}}+ \Delta ({\cal{S}}^2 - \Delta)\Big) , \non \\  f_3^{NP} &=& \frac{\mathcal{M}}{2160\Delta}\Big(- 9 {\cal{S}} + \frac{24 \mathcal{M}^2 {\cal{S}}}{\Delta^3} + \frac{28  \mathcal{M} {\cal{S}}^2}{\Delta^2} +\frac{ (-23 \mathcal{M}+ 10 {\cal{S}}^3)}{\Delta}\Big).\eea

\subsubsection{Contributions from topologically distinct graphs}

Thus we obtain the contributions from the topologically distinct graphs by adding the relevant planar and non--planar contributions. Hence we see that the contributions to the $D^8\mathcal{R}^4$ term from the topologically distinct graphs are given by

\bea \mathcal{A}^{(1)}_{D^8\mathcal{R}^4} &=&\frac{\pi^{11} \s_2^2\mathcal{R}^4}{8640\mathcal{V}_n^2} \int_0^\infty \frac{dL_1 dL_2 dL_3 }{\Delta^3} \hat{\Gamma}_{(n,n)}\Big(9  {\cal{S}}^2-8\Delta + \frac{7 \mathcal{M}^2}{\Delta^2} - \frac{12  \mathcal{M} {\cal{S}}}{\Delta}  \Big), \non \\ \mathcal{A}^{(2)}_{D^8\mathcal{R}^4} &=& \frac{\pi^{11} \s_2^2\mathcal{R}^4}{1728\mathcal{V}_n^2} \int_0^\infty \frac{dL_1 dL_2 dL_3 }{\Delta^3} \hat{\Gamma}_{(n,n)} \Big({\cal{S}}^2+ \frac{\mathcal{M}^2}{\Delta^2} -\frac{4 \mathcal{M} {\cal{S}}}{\Delta} + \Delta  \Big),\non \\  \mathcal{A}^{(3)}_{D^8\mathcal{R}^4} &=& -\frac{\pi^{11} \s_2^2\mathcal{R}^4}{864\mathcal{V}_n^2} \int_0^\infty \frac{dL_1 dL_2 dL_3 }{\Delta^3} \hat{\Gamma}_{(n,n)} \Big({\cal{S}} -\frac{2 \mathcal{M}}{\Delta}  \Big) \Big({\cal{S}}+ \frac{\mathcal{M}}{\Delta} \Big) \eea
respectively, corresponding to the graphs (i), (ii) and (iii) in figure 4 respectively. 

\begin{figure}[ht]
\begin{center}
\[
\mbox{\begin{picture}(380,100)(0,0)
\includegraphics[scale=.7]{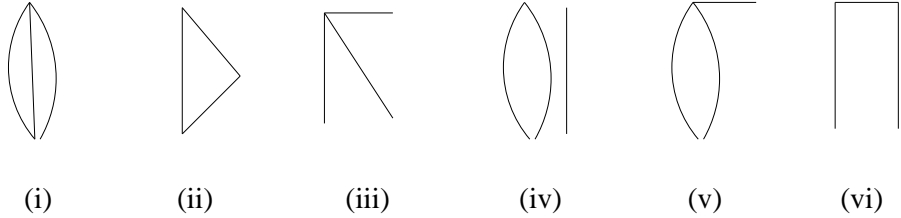}
\end{picture}}
\]
\caption{Graphs for the $D^{10}\mathcal{R}^4$ term}
\end{center}
\end{figure}

As stressed before, these are contributions from graphs which are modular invariant but not conformally invariant and are given by $Z_i$ in the conventions of~\cite{DHoker:2018mys}. The conformally and modular invariant graphs involve extra contributions and are given by $\mathcal{Z}_i$ in the conventions of~\cite{DHoker:2018mys}.   

\subsection{The $D^{10}\mathcal{R}^4$ term}

The expression \C{E3} yields several topologically distinct graphs with three links for the $D^{10} \mathcal{R}^4$ term. 
There are six such distinct graphs given in figure 5. Once again, we define $\mathcal{I} (S,T,U)$ as
\be \mathcal{I} (S,T,U) = {\mathcal{I}}_P (S,T,U) + {\mathcal{I}}_{NP} (S,T,U)\ee
as in \C{split}.

To obtain expressions for these contributions, we define the integrals
\bea g_1^i &=& \int [dt]_i \Big( \frac{5}{3}(G_{12}^3 + G_{34}^3) +\frac{1}{6} (G_{14}^3 + G_{23}^3 + G_{13}^3 + G_{24}^3) \Big), \non \\ g_2^i &=& \int [dt]_i (G_{12} G_{23} G_{31} + G_{13} G_{34} G_{41} + G_{12} G_{24} G_{41} + G_{23} G_{34} G_{42}) ,\non \\ g_3^i &=& \int [dt]_i (G_{12} G_{13} G_{14} + G_{21} G_{23} G_{24} + G_{31} G_{32}  G_{34} + G_{41} G_{42} G_{43}),\non \\ g_4^i &=& \int [dt]_i \Big(\frac{5}{3}(G_{12}^2 G_{34} + G_{12} G_{34}^2) +\frac{1}{6}(G_{14}^2 G_{23} + G_{14} G_{23}^2 + G_{13}^2 G_{24} + G_{13} G_{24}^2)\Big), \non \\ g_5^i &=& \int [dt]_i \Big( -\frac{5}{6}(G_{14} + G_{23}  + G_{13} + G_{24})(G_{12}^2 + G_{34}^2  )  \non \\ && +\frac{1}{6} (G_{12} + G_{34})(G_{14}^2 + G_{23}^2 +G_{13}^2 + G_{24}^2) -\frac{1}{3} (G_{13} + G_{24})(G_{14}^2 + G_{23}^2)\non \\ && -\frac{1}{3} (G_{14} + G_{23})(G_{13}^2 + G_{24}^2)\Big),\non \\ g_6^i &=& \int [dt]_i \Big( -\frac{5}{6}(G_{21}G_{14} G_{43} + G_{12} G_{23} G_{34} + G_{21} G_{13} G_{34} + G_{12} G_{24} G_{43}) \non \\ && +\frac{1}{6} (G_{12} + G_{34}) (G_{14} G_{23} + G_{13} G_{24})  \non \\ &&-\frac{1}{3} (G_{41} G_{13} G_{32} + G_{14} G_{42} G_{23} + G_{31}G_{14} G_{42} + G_{13} G_{32} G_{24})\Big)\eea
which arise as the integrand for the various graphs.

The analysis proceeds in a very similar way as the analysis we have done above, and we simply give the final expressions.

\subsubsection{Contributions from topologically distinct graphs}

The contributions to the $D^{10}\mathcal{R}^4$ term arising from the topologically distinct graphs (i), (ii), (iii), (iv), (v) and (vi) given in figure 5 are
\bea \mathcal{A}^{(1)}_{D^{10}\mathcal{R}^4} &=& \frac{\pi^{10} \s_2 \s_3\mathcal{R}^4}{103680\mathcal{V}_n^2} \int_0^\infty \frac{dL_1 dL_2 dL_3 }{\Delta^3} \hat{\Gamma}_{(n,n)}\Big(-\frac{18 \mathcal{M}^3}{\Delta^3} + 7 \Delta {\cal{S}} - \frac{ 20\mathcal{M}^2 {\cal{S}}}{\Delta^2} \non \\ && - \frac{3  \mathcal{M} {\cal{S}}^2}{\Delta} + 
 5 \mathcal{M} - 3 {\cal{S}}^3\Big), \non \\ 
\mathcal{A}^{(2)}_{D^{10}\mathcal{R}^4} &=&\frac{\pi^{10} \s_2 \s_3\mathcal{R}^4}{544320\mathcal{V}_n^2} \int_0^\infty \frac{dL_1 dL_2 dL_3 }{\Delta^3} \hat{\Gamma}_{(n,n)}\Big( \frac{34 \mathcal{M}^3}{\Delta^3} - 13 \Delta {\cal{S}} + \frac{138 \mathcal{M}^2 {\cal{S}}}{\Delta^2} \non \\ && + \frac{111  \mathcal{M} {\cal{S}}^2}{\Delta} - 
 13 \mathcal{M} - 33 {\cal{S}}^3\Big), \eea
\bea
\mathcal{A}^{(3)}_{D^{10}\mathcal{R}^4} &=& \frac{\pi^{10} \s_2 \s_3\mathcal{R}^4}{77760\mathcal{V}_n^2} \int_0^\infty \frac{dL_1 dL_2 dL_3 }{\Delta^3} \hat{\Gamma}_{(n,n)}\Big(-\frac{15 \mathcal{M}^3}{\Delta^3} + 3 \Delta {\cal{S}} + \frac{13  \mathcal{M}^2 {\cal{S}}}{\Delta^2} \non \\ && + \frac{12 \mathcal{M} {\cal{S}}^2}{\Delta} + 
 3  (\mathcal{M} - 2 {\cal{S}}^3)\Big), \non \\ 
\mathcal{A}^{(4)}_{D^{10}\mathcal{R}^4} &=& \frac{\pi^{10} \s_2 \s_3\mathcal{R}^4}{311040\mathcal{V}_n^2} \int_0^\infty \frac{dL_1 dL_2 dL_3 }{\Delta^3} \hat{\Gamma}_{(n,n)}\Big(\frac{31 \mathcal{M}^3}{\Delta^3} - 13 \Delta {\cal{S}} + \frac{21  \mathcal{M}^2 {\cal{S}}}{\Delta^2} \non \\ && + \frac{30 \mathcal{M} {\cal{S}}^2 }{\Delta}- 
  19 \mathcal{M} - 6 {\cal{S}}^3\Big), \non \\ 
\mathcal{A}^{(5)}_{D^{10}\mathcal{R}^4} &=& \frac{\pi^{10} \s_2 \s_3\mathcal{R}^4}{1088640\mathcal{V}_n^2} \int_0^\infty \frac{dL_1 dL_2 dL_3 }{\Delta^3} \hat{\Gamma}_{(n,n)}\Big(\frac{64 \mathcal{M}^3}{\Delta^3} - 253 \Delta {\cal{S}} + \frac{156  \mathcal{M}^2 {\cal{S}}}{\Delta^2} \non \\ && - \frac{384  \mathcal{M} {\cal{S}}^2}{\Delta} + 
 5  (109 \mathcal{M} + 36 {\cal{S}}^3)\Big), \non \\ 
\mathcal{A}^{(6)}_{D^{10}\mathcal{R}^4} &=& \frac{\pi^{10} \s_2 \s_3\mathcal{R}^4}{241920\mathcal{V}_n^2} \int_0^\infty \frac{dL_1 dL_2 dL_3 }{\Delta^3} \hat{\Gamma}_{(n,n)}\Big(\frac{3 \mathcal{M}^3}{\Delta^3} + 32 \Delta {\cal{S}} - \frac{29  \mathcal{M}^2 {\cal{S}} }{\Delta^2} \non \\ && - \frac{81  \mathcal{M} {\cal{S}}^2}{\Delta} 
 -66 \mathcal{M} + 15 {\cal{S}}^3\Big)\eea
respectively. Thus from these simple example we see how the worldline formalism gives the contributions from the individual graphs in a very natural way.   

\section{Matching with the structure of the genus two type II string amplitude}

The above analysis gives us the contributions from the topologically distinct graphs to the low momentum expansion of the supergravity amplitude up to the $D^{10} \mathcal{R}^4$ term. This must match with the field theory limit of the contributions obtained from the low momentum expansion of the string amplitude, as we now explain.

Consider the four graviton amplitude at genus two in type II string theory compactified on $T^n$, where
$G_{IJ}$ is the metric on $T^n$, and $B_{IJ}$ are the components of the NS--NS two form along $T^n$. The low momentum expansion of this amplitude yields an expansion in powers of $\alpha'$. Every term in the expansion is an expression of the form  
\be \label{2loop}(e^{-2\phi} \mathcal{V}_n)^{-1} \int_{\mathcal{F}_2} d\mu_2 f(\Omega, \bar\Omega) Z_{lat} (G_{IJ},B_{IJ}; \Omega,\bar\Omega), \ee
where $\mathcal{F}_2$ is the fundamental domain of $Sp(4,\mathbb{Z})$ whose explicit details are not relevant for our purposes. Here $\phi$ is the dilaton, and the $Sp(4,\mathbb{Z})$ invariant measure $d\mu_2$ is given by
\be d\mu_2 = \frac{1}{({\rm det} Y)^3} \prod_{\a \leq \b} i d\Omega_{\a\b} \wedge d\bar\Omega_{\a\b},\ee
where $\Omega_{\a\b}$ ($\a,\b =1,2$) is the period matrix, and $\Omega = X + iY$, where $X$ and $Y$ are matrices with real entries. The lattice factor is given by 
\be \label{Z} Z_{lat} (G_{IJ},B_{IJ}; \Omega,\bar\Omega) = \mathcal{V}_n^2 \sum_{m^I_\a, n^I_a \in \mathbb{Z}}e^{-\pi(G+B)_{IJ}(m^I_\a + \Omega_{\a\g} n^I_\g)Y^{-1}_{\a\b}(m^J_\b + \bar\Omega_{\b\d} n^J_\d)},\ee
where $(Y^{-1})_{IJ} \equiv Y^{-1}_{IJ}$. 

Consider the contribution obtained from \C{2loop} in the field theory limit which is the complete non--separating degeneration limit of the genus two Riemann surface. In this limit, the genus two Riemann surface degenerates into the two loop skeleton diagram where $L_i$ is large, and one can make a systematic $1/L_i$ expansion\footnote{Essentially on the Riemann surface, this corresponds to parametrizing a tube locally as $q= e^{2\pi i \tau}$, and taking ${\rm Im}\tau \rightarrow \infty$, and performing the angular integral over $\tau_1$. In the field theory limit, ${\rm Im} \tau$ is $L$ where $L$ is the proper time in the Schwinger representation of the propagator.}. Thus the field theory analysis must reproduce this answer at leading order in the large $L_i$ expansion.

We shall not be concerned with overall factors, and consider the leading contribution in this limit\footnote{Note that the supergravity calculations are done in the Einstein frame, and need to be converted to the string frame to match with the worldsheet calculations, hence differing by a dilaton dependent overall factor.}. We have that  
\be d\mu_2 \rightarrow \frac{dL_1 dL_2 dL_3}{\Delta^3}, \ee
where
\be Y \rightarrow  \left( \begin{array}{cc} L_1 + L_2  & -L_2\\ -L_2 & L_2 + L_3\end{array}\right).\ee
Thus \C{inv} in the worldline formulation is indeed $Y^{-1}$.

Also to match with the supergravity calculations, we keep only the momentum modes and neglect the winding modes (thus $n^I_\a =0$ in \C{Z}). We also set $B_{IJ} =0$, giving us
\be Z_{lat} (G_{IJ},B_{IJ}; \Omega,\bar\Omega) \rightarrow \hat\Gamma_{(n,n)}.\ee 
Thus the measure for the toroidally compactified theory is universal and is given by
\be \frac{dL_1 dL_2 dL_3}{\Delta^3} \hat\Gamma_{(n,n)} \ee
which is then integrated over $L_i$. This precisely reproduces the supergravity calculations. Thus $f(\Omega,\bar\Omega)$ in the field theory limit must reduce to the integrand in the supergravity analysis which is then integrated with the measure mentioned above.  

We now consider the explicit expression for the four graviton amplitude in the toroidally compactified type II theory. We have that~\cite{D'Hoker:2005vch}
\be \label{string}\mathcal{A}_{string} = \frac{\pi}{64}  (e^{-2\phi}\mathcal{V}_n)^{-1} \mathcal{R}^4\int_{\mathcal{F}_2} d\mu_2  Z_{lat} (G_{IJ},B_{IJ}; \Omega,\bar\Omega) F (\Omega,\bar\Omega)\ee
where $F(\Omega,\bar\Omega)$ involves the Koba--Nielsen factor and is given by 
\be \label{F}F (\Omega,\bar\Omega) = \int_{\S^4} \frac{\vert \mathcal{W} \vert^2}{({\rm det}Y)^2} e^{-\sum_{r\neq s =1}^4 p_r \cdot p_s G(z_r,z_s) }.\ee
where $\Sigma^4$ refers to an integral over the positions of insertion of the four vertex operators over the Riemann surface. 
In \C{F}, we also have that
\be 3 \mathcal{W} = (T-U)\Delta(1,2)  \Delta(3,4) + (S-T) \Delta(1,3) \Delta(4,2)+ (U-S) \Delta(1,4)\Delta(2,3),\ee
where
\be \Delta(i,j) = \epsilon^{\a\b} \omega_\a (z_i) \omega_\b (z_j),\ee
where $\omega_\a (z)$ are the abelian differentials.  

From the low momentum expansion of \C{string}, we easily see that precisely the graphs that arise in the supergravity analysis are the ones that arise in the string amplitude\footnote{That the field theory limit of the modular graph for the $D^6\mathcal{R}^4$ term in the string amplitude reproduces the supergravity calculation has been checked in~\cite{D'Hoker:2014gfa}.}. This is what is expected, and the string theory graphs are the $Sp(4,\mathbb{Z})$ completions of the graphs in supergravity. It will be interesting using the field theory analysis to obtain constraints on the structure of these the $Sp(4,\mathbb{Z})$ invariant graphs, some of which have been studied in~\cite{D'Hoker:2005jhf,D'Hoker:2013eea,D'Hoker:2014gfa,Basu:2015dqa,Pioline:2015nfa,DHoker:2017pvk}. 

\section{Eigenvalue equations satisfied by the supergravity graphs}

Thus we have obtained explicit expressions for the topologically distinct graphs that contribute to terms in the low momentum expansion of the supergravity amplitude. What remains is to perform the integral over the moduli  space characterized by $L_i$. To do so, it is very useful to make a change of variables
\be \tau_1 = \frac{L_1}{L_1 + L_2}, \quad \tau_2 = \frac{\sqrt{\Delta}}{L_1 + L_2}, \quad V = \frac{1}{\sqrt{\Delta}}\ee
where $V$ and $\tau$ parametrize the volume and complex structure modulus of an auxiliary torus respectively. While the $V$ dependence of the integrand is easily fixed by scaling arguments, the $\tau$ dependence is involved. For the terms that arise in the low momentum expansion of \C{E1} which include the contributions from all the graphs put together, this has been obtained in~\cite{Green:1999pu,Green:2005ba,Green:2008bf}, to which we refer the reader. The main observation is that the integrand satisfies an intricate pattern of Laplace eigenvalue equations with source terms, where the $SL(2,\mathbb{Z})$ invariant Laplacian is given by   
\be \Delta_\tau = 4\tau_2^2\frac{\p^2}{\p\tau \p\bar\tau}.\ee

The analysis for the contribution from topologically distinct graphs that result from the low momentum expansion of \C{E3} follow along same lines, and we mention only the results. Once again we see how the worldline formalism gives the separate contributions in a systematic way. For simplicity we obtain eigenvalue equations up to the $D^8\mathcal{R}^4$ term, while the terms at higher orders in the derivative expansion can be analyzed in the same way. 

For the $D^4\mathcal{R}^4$ term, we have that
\be \mathcal{A}_{D^4\mathcal{R}^4} = \frac{\pi^{13}\s_2 \mathcal{R}^4}{\mathcal{V}_n^2} \int_0^\infty dV V^2 \int_{\mathcal{F}_1} \frac{d^2\tau}{\tau_2^2} \hat\Gamma_{(n,n)}\ee
where we have integrated with a modular invariant measure over $\mathcal{F}_1$, the fundamental domain of $SL(2,\mathbb{Z})$, and
 \be \hat\Gamma_{(n,n)} = \mathcal{V}_n^2 \sum e^{-\pi G_{IJ}V(m+ n\tau)_I(m+n\bar\tau)_J/\tau_2 }.\ee

For terms at higher orders in the momentum expansion, it is useful to define
\be T = -\tau_1^2 + \vert \tau_1 \vert\ee
and the integrand is no more  modular invariant. For the $D^6\mathcal{R}^4$ term, we get
\be \mathcal{A}_{D^6\mathcal{R}^4} = \frac{\pi^{12}\s_2 \mathcal{R}^4}{12\mathcal{V}_n^2} \int_0^\infty dV V \int_{\mathcal{F}_1} \frac{d^2\tau}{\tau_2^2} \hat\Gamma_{(n,n)} \Big(\tau_2 + \frac{1-6T}{\tau_2} + \frac{5T^2}{\tau_2^3}\Big)\ee
where the integrand satisfies the eigenvalue equation
\be (\Delta_\tau -12) \Big(\tau_2 + \frac{1-6T}{\tau_2} + \frac{5T^2}{\tau_2^3}\Big) = -12 \tau_2 \delta (\tau_1).\ee

For the $D^8\mathcal{R}^4$ term, we define
\bea  \mathcal{A}^{(1)}_{D^8\mathcal{R}^4} &=& \frac{\pi^{11}\s_2^2 \mathcal{R}^4}{1440\mathcal{V}_n^2}\int_0^\infty dV  \int_{\mathcal{F}_1} \frac{d^2\tau}{\tau_2^2} \hat\Gamma_{(n,n)}\mathcal{B}_1, \non \\ 
\mathcal{A}^{(2)}_{D^8\mathcal{R}^4} &=& \frac{\pi^{11}\s_2^2 \mathcal{R}^4}{288\mathcal{V}_n^2}\int_0^\infty dV  \int_{\mathcal{F}_1} \frac{d^2\tau}{\tau_2^2} \hat\Gamma_{(n,n)}\mathcal{B}_2, \non \\ 
\mathcal{A}^{(3)}_{D^8\mathcal{R}^4} &=& \frac{\pi^{11}\s_2^2 \mathcal{R}^4}{144\mathcal{V}_n^2}\int_0^\infty dV  \int_{\mathcal{F}_1} \frac{d^2\tau}{\tau_2^2} \hat\Gamma_{(n,n)}\mathcal{B}_3,\eea
where $\mathcal{B}_i$ is only dependent on $\tau$, and is given by
\bea \mathcal{B}_1 &=&  \frac{7 T^4}{\tau_2^6} + \frac{2 (6 - 13 T) T^2}{\tau_2^4} + \frac{9 + 10 T (4 T-3)}{\tau_2^2} + 10 (1 - 3 T)+ 9 \tau_2^2, \non \\ \mathcal{B}_2 &=&  \frac{T^4}{\tau_2^6} + \frac{2 (2 - 3 T) T^2}{\tau_2^4} +\frac{1 + 2 T ( 5 T-3)}{\tau_2^2} + 3 (1 - 2 T)+ \tau_2^2, \non \\ \mathcal{B}_3 &=&   \frac{2 T^4}{\tau_2^6} - \frac{T^2(1+3T)}{\tau_2^4} - \frac{(1-3T + T^2)}{\tau_2^2} - 2+ 3 T- \tau_2^2 .\eea
Now each $\mathcal{B}_i$ can be written as a sum over eigenfunctions of the $SL(2,\mathbb{Z})$ invariant Laplacian with different eigenvalues. The relevant eigenfunctions and eigenvalues are:

(i) \be (\Delta_\tau  - 42 ) \mathcal{C} = -40 \delta(\tau_1) \tau_2 (\tau_2 +\tau_2^{-1}),\ee
\be \mathcal{C} = \frac{33T^4}{\tau_2^6} +\frac{6T^2 (3-14 T)}{\tau_2^4} + \frac{1-20 T + 70 T^2}{\tau_2^2} + \frac{10(3-14 T)}{7} +\tau_2^2,\ee

(ii) \be  (\Delta_\tau  - 20 ) \mathcal{D} = 18 \delta(\tau_1) \tau_2 (\tau_2 +\tau_2^{-1}),\ee
\be \mathcal{D} = \frac{7T^2(T-1)}{\tau_2^4} + \frac{9T - 15 T^2 -1}{\tau_2^2} +\frac{3(15T-4)}{5}-\tau_2^2,\ee

(iii)
\be (\Delta_\tau -6) \mathcal{E} = -4 \delta(\tau_1) \tau_2 (\tau_2 +\tau_2^{-1}),\ee
\be \mathcal{E} = \frac{(T-1)^2}{\tau_2^2} + 1-2T +\tau_2^2.\ee
This leads to
\bea  \mathcal{B}_1 &=& \frac{7\mathcal{C}}{33}  -\frac{90\mathcal{D}}{77}+\frac{160\mathcal{E}}{21} -\frac{4}{3} , \non \\   \mathcal{B}_2  &=& \frac{\mathcal{C}}{33} -\frac{38\mathcal{D}}{77}  +\frac{10\mathcal{E}}{21} +\frac{127}{105}, \non \\  \mathcal{B}_3 &=& \frac{2\mathcal{C}}{33} +\frac{23\mathcal{D}}{77} -\frac{16\mathcal{E}}{21} - \frac{82}{105}.\eea
Then the integrals can be performed along the lines of~\cite{Green:2008bf} in a straightforward manner. 

Thus using the worldline formalism we have studied very simple cases where the contributions from topologically distinct graphs can be analyzed separately, which should impose useful constraints on the general structure of scattering amplitudes.   


\providecommand{\href}[2]{#2}\begingroup\raggedright\endgroup

\end{document}